\documentclass{pasj00}
\begin{document}
\SetRunningHead
{J.\ Fukue and C.\ Akizuki}
{Relativistic Radiation Hydrodynamical Accretion Disk Winds}
\Received{yyyy/mm/dd}
\Accepted{yyyy/mm/dd}

\title{Relativistic Radiation Hydrodynamical Accretion Disk Winds}

\author{Jun \textsc{Fukue} and Chizuru \textsc{Akizuki}
\thanks{Present address: Center for Computational Physics, 
University of Tsukuba, Tennoudai 1-1-1, Tsukuba, Ibaraki, 305-8577} } 
\affil{Astronomical Institute, Osaka Kyoiku University, 
Asahigaoka, Kashiwara, Osaka 582-8582}
\email{fukue@cc.osaka-kyoiku.ac.jp}



\maketitle


\begin{abstract}
Accretion disk winds browing off perpendicular to a luminous disk
are examined in the framework of fully special
relativistic radiation hydrodynamics.
The wind is assumed to be steady, vertical, and isothermal.
Using a velocity-dependent variable Eddington factor,
we can solve the rigorous equations of relativistic radiative hydrodynamics,
and can obtain radiatively driven winds
accelerated up to the {\it relativistic} speed.
%
For less luminous cases,
disk winds are transonic types passing through saddle type critical points,
and the final speed of winds increases
as the disk flux and/or the isothermal sound speed increase.
For luminous cases, on the other hand,
disk winds are always supersonic,
since critical points disappear
due to the characteristic nature of the disk gravitational fields.
The boundary between the transonic and supersonic types is located
at around 
$\hat{F}_{\rm c} \sim 0.1 (\varepsilon+p)/(\rho c^2)/\gamma_{\rm c}$,
where $\hat{F}_{\rm c}$ is the radiative flux at the critical point
normalized by the local Eddington luminosity,
$(\varepsilon+p)/(\rho c^2)$ is the enthalpy of the gas
divided by the rest mass energy,
and $\gamma_{\rm c}$ is the Lorentz factor of the wind velocity
at the critical point.
In the transonic winds,
the final speed becomes 0.4--0.8$c$ for typical parameters,
while it can reach $\sim c$ in the supersonic winds.
\end{abstract}

\section{Introduction}

Mass outflows -- accretion disk winds -- from a luminous disk
are observed in various active objects,
such as cataclysmic variables (CVs),
supersoft X-ray sources (SSXSs),
microquasars ($\mu$QSOs),
broad absorption line quasars (BAL QSOs), and so on.
Radiatively driven wind emanating from an accretion disk
is a clue to the formation mechanism of
astrophysical jets and winds in these objects
(see Kato et al. 1998, 2007 for a review of accretion disks).

So far, radiatively driven outflows from a luminous disk
have been extensively studied by many researchers
(Bisnovatyi-Kogan, Blinnikov 1977; Katz 1980; Icke 1980; Melia, K\"onigl 1989; 
Misra, Melia 1993; Tajima, Fukue 1996, 1998; Watarai, Fukue 1999;
Hirai, Fukue 2001; Fukue et al. 2001; Orihara, Fukue 2003),
and by numerical simulations
(Eggum et al. 1985, 1988; Kley 1989; Okuda et al. 1997;
Kley, Lin 1999; Okuda, Fujita 2000;
Okuda 2002; Okuda et al. 2005; Ohsuga et al. 2005; Ohsuga 2006).
In almost all of these studies, however,
the luminous disk was treated as an external radiation source
(optically thin approximation),
and the radiation transfer in the flow was not solved.
Although radiation hydrodynamical equations were solved
in the numerical simulations,
the methods were rather limited; e.g.,
the equilibrium between gas and radiation was assumed,
the flux-limited diffusion approximation was adopted,
the flow velocity was subrelativistic on the order of $0.1 c$, 
or the optically thick to thin transition was not properly treated.
Up to now, in relation to accretion disk winds
no one solved the fully relativistic radiation hydrodynamical equations.


Recently, 
radiation hydrodynamical mass outflows have been examined
for the first time in the fully relativistic cases 
for the plane-parallel case (Fukue 2005, 2006; Fukue, Akizuki 2006;
see also Akizuki and Fukue 2007 for the spherical case).
In these studies, however, the gas pressure was ignored for simplicity,
although the relativistic radiation hydrodynamical equations were solved.
In this paper, we thus take into account the gas pressure,
and obtain the relativistic radiation hydrodynamical {\it winds}
browing off from a luminous flat disk
within the framework of a fully special relativistic regime.

In the next section
we describe the basic equations in the vertical direction,
and examine critical points.
In section 3
we obtain transonic winds as well as supersonic ones.
The final section is devoted to concluding remarks.


\section{Basic Equations and Boundary Conditions}

Let us suppose a luminous flat disk, inside of which
gravitational or nuclear energy is released
via viscous heating or other processes.
The radiation energy is transported in the vertical direction,
and the disk gas, itself, also moves in the vertical direction
due to the action of radiation pressure
(steady plane-parallel approximation).
We do not consider the rotation of the gas.
As for the order of the flow velocity $v$,
we consider the fully relativistic regime,
where the terms are retained up to the second order of $(v/c)$.
As for the gravity, on the other hand,
we adopt the pseudo-Newtonian approximation
(Paczy\'nski, Wiita 1980),
since we do not consider the region very close 
to the Schwarzschild radius.
For simplicity, in the present paper,
we assume that the gas is {\it isothermal},
because we focus our attention on the general properties
of transonic disk winds driven by disk radiation fields
under relativistic radiation hydrodynamics.
We further assume the radiative equilibrium and
use the gray approximation.
Finally, in order to close moment equations,
we adopt the {\it velocity-dependent variable Eddington factor}
proposed by Fukue (2006).

\subsection{Basic Equations}

Under these assumptions,
the radiation hydrodynamic equations
for steady vertical ($z$) winds are described as follows
(Kato et al. 1998, 2007; Fukue 2006).

The continuity equation is
\begin{equation}
   \rho cu = J ~(={\rm const.}),
\label{rho1}
\end{equation}
where $\rho$ is the proper gas density, $u$ the vertical four velocity, 
$J$ the mass-loss rate per unit area,
and $c$ the speed of light.
The four velocity $u$ is related to the proper three velocity $v$ by
$u=\gamma v/c$, where $\gamma$ is the Lorentz factor,
$\gamma=\sqrt{1+u^2}=1/\sqrt{1-(v/c)^2}$.

The equation of motion is, within the present approximation,
\begin{eqnarray}
   c^2u\frac{du}{dz} &=& -\frac{GMz}{(R-r_{\rm g})^2 R}
                         -\gamma^2 \frac{c^2}{\varepsilon+p}\frac{dp}{dz}
\nonumber \\
         &&       +\frac{\rho c^2}{\varepsilon+p}
                   \frac{\kappa_{\rm abs}+\kappa_{\rm sca}}{c}
\nonumber \\
         &&
            \times  \left[ F \gamma (1+2u^2) - (cE+cP)\gamma^2 u \right],
\label{u1}
\end{eqnarray}
where $M$ is the mass of the central object,
$R$ $=\sqrt{r^2+z^2}$, $r$ being the radius,
$r_{\rm g}$ ($=2GM/c^2$) the Schwarzschild radius,
$\varepsilon$ the gas internal energy per unit proper volume,
$p$ the gas pressure measured in the comoving frame,
$\kappa_{\rm abs}$ and $\kappa_{\rm sca}$
the absorption and scattering opacities (gray),
defined in the comoving frame,
$E$ the radiation energy density, $F$ the radiative flux, and
$P$ the radiation pressure observed in the inertial frame
(Fukue 2006; Kato et al. 2007).
The first term in the square bracket on the right-hand side
of equation (\ref{u1}) means the radiatively-driven force,
which is modified to the order of $u^2$, whereas
the second term is the radiation drag force,
which is also modified, but roughly proportional to the velocity.
Compared with the previous researches (Fukue 2006; Fukue and Akizuki 2006),
we have added the pressure gradient force,
the second term on the right-hand side of equation (\ref{u1}),
and the related factor, $\rho c^2/(\varepsilon+p)$
to the third term.

The radiative equilibrium condition is,
in the inertial frame, written as
\begin{equation}
   0 =  j - \kappa_{\rm abs} cE \gamma^2 
                  - \kappa_{\rm abs} cP u^2
                  + 2 \kappa_{\rm abs} F \gamma u,
\label{j1}
\end{equation}
where $j$ is the emissivity defined in the comoving frame
(Fukue 2006; Kato et al. 2007).

For radiation fields, the zeroth-moment equation becomes
\begin{eqnarray}
   \frac{dF}{dz} &=& \rho \gamma
         \left[ j - \kappa_{\rm abs} cE
                 + \kappa_{\rm sca}(cE+cP)u^2  \right.
\nonumber \\
    &&   \left. + \kappa_{\rm abs}Fu/\gamma
               -\kappa_{\rm sca}F ( 1+v^2/c^2 )\gamma u \right]
\nonumber \\
    &=& -\rho (\kappa_{\rm abs}+\kappa_{\rm sca}) \frac{u}{\gamma}
\nonumber \\
    && \times    \left[ F(1+2u^2)\gamma - (cE+cP)\gamma^2 u \right],
\label{F1}
\end{eqnarray}
where we use equation (\ref{j1}) for the second equality.
The first-moment equation is
\begin{eqnarray}
   \frac{dP}{dz} &=& \frac{\rho \gamma}{c} 
         \left[ j(u/\gamma) - \kappa_{\rm abs} F
                  + \kappa_{\rm abs}cP (u/\gamma) \right.
\nonumber \\
     && \left. -\kappa_{\rm sca}F(1+2u^2)
               +\kappa_{\rm sca}(cE+cP)\gamma u \right]
\nonumber \\
     &=& -\rho (\kappa_{\rm abs}+\kappa_{\rm sca}) \frac{1}{c}
\nonumber \\
    && \times    \left[ F(1+2u^2)\gamma - (cE+cP)\gamma^2 u \right],
\label{P1}
\end{eqnarray}
where we use equation (\ref{j1}) for the second equality
(Fukue 2006; Kato et al. 2007).

In order to close moment equations for radiation fields,
we need some closure relation.
Instead of the usual Eddington approximation,
we here adopt a {\it velocity-dependent} 
variable Eddington factor $f(\beta)$,
\begin{equation}
   P_0 = f(\beta) E_0
\label{close0}
\end{equation}
in the comoving frame,
where $P_0$ and $E_0$ are the quantities in the comoving frame,
and $\beta=v/c$.
If we adopt this form (\ref{close0}) as the closure relation
in the comoving frame,
the transformed closure relation in the inertial frame is
\begin{equation}
   cP \left( 1 + u^2 - fu^2 \right) = 
   cE \left( f\gamma^2 - u^2 \right) 
   + 2 F \gamma u \left( 1 - f \right),
\label{close}
\end{equation}
or equivalently,
\begin{equation}
   cP \left( 1 - f\beta^2 \right) =
   cE \left( f - \beta^2 \right) + 2F\beta \left( 1 - f \right).
\label{close_beta}
\end{equation}
As a form of the function $f(\beta)$,
we adopt the simplest one:
\begin{equation}
   f(\beta) = \frac{1}{3} + \frac{2}{3} \beta
\label{Eddington}
\end{equation}
for a plane-parallel geometry (Fukue 2006; Fukue, Akizuki 2006;
cf. Akizuki and Fukue 2007 for a spherically symmetric geometry).

Using continuity equation (\ref{rho1})
and equation of motion (\ref{u1}), after some manipulations,
we obtain the so-called {\it wind} equation,
\begin{eqnarray}
    \frac{du}{dz}
          &=& \frac{u}{\gamma^2 \left( v^2 - c_{\rm T}^2 \right)} 
              \left\{ -\frac{GMz}{(R-r_{\rm g})^2 R} \right.
\nonumber \\
         && +  \frac{\rho c^2}{\varepsilon+p}
                   \frac{\kappa_{\rm abs}+\kappa_{\rm sca}}{c}
\nonumber \\
        && \times \left. \left[ F \gamma (1+2u^2) 
                - \!\!\!\!\!\!\! \frac{~}{~}
               (cE+cP)\gamma^2 u \right]
               \right\},
\label{u2}
\end{eqnarray}
or equivalently,
\begin{eqnarray}
    \frac{dv}{dz}
          &=& \frac{v}{\gamma^4 \left( v^2 - c_{\rm T}^2 \right)} 
              \left\{ -\frac{GMz}{(R-r_{\rm g})^2 R} \right.
\nonumber \\
         && +      \delta^2
                   \frac{\kappa_{\rm abs}+\kappa_{\rm sca}}{c}
\nonumber \\
         && \left. \times \gamma^3 \left[ F \gamma 
                 \left( 1+\frac{v^2}{c^2} \right)
                - (cE+cP)\frac{v}{c} \right] \right\}.
\label{v2}
\end{eqnarray}
Here, $c_{\rm T}$ is the constant isothermal sound speed,
defined by 
\begin{equation}
   c_{\rm T}^2 \equiv \frac{c^2p}{\varepsilon+p} ~~~({\rm const.}).
\end{equation}
Under the present isothermal assumption,
the factor, 
\begin{equation}
    \delta^2 \equiv \frac{\rho c^2}{\varepsilon+p} ~~~({\rm const.}),
\end{equation}
is also a constant parameter.

Eliminating $\rho$ and $E$ 
with the help of equations (\ref{rho1}) and (\ref{close}),
equations (\ref{v2}), (\ref{F1}), and (\ref{P1}) become
\begin{eqnarray}
   \frac{dv}{dz}
          &=& \frac{v}{\gamma^4 \left( v^2 - c_{\rm T}^2 \right) }
              \left[ -\frac{GMz}{(R-r_{\rm g})^2 R} \right.
\nonumber \\
         &&+       \delta^2
                   \frac{\kappa_{\rm abs}+\kappa_{\rm sca}}{c}
\nonumber \\
         && \times \left. \gamma \frac{F(f+\beta^2) -cP(1+f)\beta}{f-\beta^2} \right],
\label{v3}
\\
   \frac{dF}{dz} &=& 
    - J \frac{\kappa_{\rm abs}+\kappa_{\rm sca}}{c} 
                    \frac{F(f+\beta^2) -cP(1+f)\beta}{f-\beta^2},
\label{F3}
\\
   c\frac{dP}{dz} &=& 
     - \frac{J}{u} \frac{\kappa_{\rm abs}+\kappa_{\rm sca}}{c} \gamma 
                    \frac{F(f+\beta^2) -cP(1+f)\beta}{f-\beta^2}.
\label{P3}
\end{eqnarray}
For the optical depth $\tau$,
defined by $d\tau \equiv -(\kappa_{\rm abs}+\kappa_{\rm sca})\rho dz$,
we have
\begin{equation}
   \frac{d\tau}{dz} = 
    - J \frac{\kappa_{\rm abs}+\kappa_{\rm sca}}{c} \frac{1}{u}.
\label{tau3}
\end{equation}

Finally, using the nondimensional variables, such as
$\hat{z}=z/r_{\rm g}$,
$\beta = v/c$,
$\hat{F}=F/[L_{\rm E}/(4\pi r_{\rm g}^2)]$,
$\hat{P}=cP/[L_{\rm E}/(4\pi r_{\rm g}^2)]$,
where $L_{\rm E}$ [$=4\pi cGM/(\kappa_{\rm abs}+\kappa_{\rm sca})$]
is the Eddington luminosity,
basic equations (\ref{v3})--(\ref{tau3}) are normalized as
\begin{eqnarray}
\!\!\!\!\!
\!\!\!\!\!
    \frac{d\beta}{dz}
          &=& \frac{\beta}{\gamma^4 \left( \beta^2 - \hat{c}_{\rm T}^2 \right)}
\nonumber \\
\!\!\!\!\!
\!\!\!\!\!
          &\times& \left[ -\frac{\hat{z}}{2(\hat{R}-1)^2 \hat{R}}
         +     \frac{\delta^2 \gamma }{2} 
\frac{\hat{F}(f+\beta^2) -\hat{P}(1+f)\beta}{f-\beta^2}
        \right],
\label{v4}
\\
\!\!\!\!\!
\!\!\!\!\!
   \frac{d\hat{F}}{d\hat{z}} &= &
    - \frac{\hat{J}}{2}  
\frac{\hat{F}(f+\beta^2) -\hat{P}(1+f)\beta}{f-\beta^2},
\label{F4}
\\
\!\!\!\!\!
\!\!\!\!\!
   \frac{d\hat{P}}{d\hat{z}} &= &
     - \frac{\hat{J}}{2\beta} 
\frac{\hat{F}(f+\beta^2) -\hat{P}(1+f)\beta}{f-\beta^2}\label{P4}
\\
   \frac{d\tau}{d\hat{z}} &= &
    - \frac{\hat{J}}{2\gamma\beta},
\label{tau4}
\end{eqnarray}
where $\hat{c}_{\rm T}^2 = c_{\rm T}^2/c^2 = p/(\varepsilon+p)$, 
$\delta^2=\rho c^2/(\varepsilon+p)$, and
$\hat{J}=c^2 J/[L_{\rm E}/(4\pi r_{\rm g}^2)]$.

We solve equations (\ref{v4})--(\ref{tau4})
under regularity conditions at the sonic point and
appropriate boundary conditions at the moving surface
with a variable Eddington factor (\ref{Eddington}).

\subsection{Regularity Conditions and Boundary Conditions}

As is well known,
equation (\ref{v4}) has critical points,
where the denominator and numerator vanish, simultaneously.
Hence, transonic solutions, which passes through critical points,
satisfy the {\it regularity conditions} at the critical point.
In order to obtain transonic wind solutions, in this paper,
we first search the location of critical points,
examine the types of critical points, and
calculate the transonic solutions from the critical point
both inward and outward directions.

In contrast to a simple spherical flow
under the gravity of the central object,
the gravitational field of the disk wind
does not monotonically decrease, but
increases at first and then decreases
(Fukue 2002).
As a result, there may appear multiple critical points.
In figure 1, we show some typical examples.

\begin{figure}
  \begin{center}
  \FigureFile(80mm,80mm){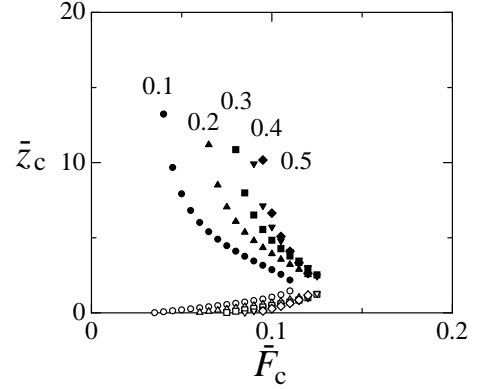}
  \end{center}
\caption{
Loci and types of critical points
as a function of the radiative flux $\hat{F}_{\rm c}$ at the critical point
for several values of the isothermal sound speed $\hat{c}_{\rm T}$.
Filled symbols represent the saddle type,
whereas open symbols denote the center type.
The values of $\hat{c}_{\rm T}$ are attached on each symbols.
The other parameters are fixed as
$\delta=1$, $\tau_{\rm c}=1$, and $\hat{P}_{\rm c}=0.1$.
}
\end{figure}

In figure 1,
the height $\hat{z}_{\rm c}$ of critical points are plotted
as a function of the radiative flux $\hat{F}_{\rm c}$ at the critical point
for several values of the isothermal sound speed $\hat{c}_{\rm T}$.
Filled symbols represent the saddle type,
through which physical solutions can path,
whereas open symbols denote the center type,
where no solution can path
(Kato et al. 2007).
The values of $\hat{c}_{\rm T}$ are attached on each symbols.
The other parameters are fixed as
$\delta=1$, $\tau_{\rm c}=1$, and $\hat{P}_{\rm c}=0.1$.

As is easily seen in figure 1,
in less-luminous cases
there appear two positions for the same values of $\hat{F}_{\rm c}$.
Of these,
the upper point is usually a saddle type,
whereas the lower one is a center type.
On the other hand,
in luminous cases
there is no critical point.
Roughly speaking,
two critical points exist when
$\delta^2 \gamma_{\rm c} \hat{F}_{\rm c} < 0.1 $, or
\begin{equation}
   \frac{F_{\rm c}}{L_{\rm E}/(4\pi r_{\rm g}^2)} 
   < 0.1 \frac{\varepsilon+p}{\rho c^2} \frac{1}{\gamma_{\rm c}},
\end{equation}
where ${F}_{\rm c}$ is the radiative flux at the critical point,
$(\varepsilon+p)/(\rho c^2)$ is the enthalpy of the gas
divided by the rest mass energy, and
$\gamma_{\rm c}$ is the Lorentz factor of the wind velocity
at the critical point.

In addition to the above regularity conditions,
at the wind top of $\tau=0$
we must impose the boundary conditions,
which is different from the usual ones for a static photosphere,
since the wind top moves upward at a relativistic speed
(Fukue 2005).
In the case of a ``moving photosphere'',
due to relativistic aberration and Doppler effect
(cf. Fukue 2000),
the boundary conditions imposed for the radiation quantities become
\begin{equation}
   \frac{cP_{\rm s}}{F_{\rm s}}
   = \frac{2+6\beta_{\rm s}+6\beta_{\rm s}^2}
       {3+8\beta_{\rm s}+3\beta_{\rm s}^2},
\label{bc2}
\end{equation}
where the subscript `s' denotes the quantites at the wind top,
and $\beta_{\rm s}$ is a final speed at the wind top.

\section{Relativistic Accretion Disk Winds}

In this section
we show typical solutions of
relativistic radiation hydrodynamical accretion disk winds
for transonic and supersonic cases.

In order to obtain a transonic solution,
by numerically solving equations (\ref{v4})--(\ref{tau4}),
we must specify the initial values of $r$,
and $z_{\rm c}$, $\beta_{\rm c}$, $F_{\rm c}$, $P_{\rm c}$, $\tau_{\rm c}$
at the critical point
for a given set of parameters of
$c_{\rm T}$, $\delta$, and $J$.
However, there are three restrictive conditions;
two regularity conditions at the critical points, and
one boundary condition at the wind top.
Hence,
we have six freedoms.
In order to see the effect of the radiative force
and the gas pressure,
in the present paper
we set $\hat{r}=3$, $\tau_{\rm c}=1$,
$\delta=1$, and $\hat{J}=0.1$.
If we give the values of $F_{\rm c}$ and $c_{\rm T}$,
the values of $z_{\rm c}$, $\beta_{\rm c}$, and $P_{\rm c}$
are automatically determined.

Several examples of transonic solutions
are shown in figure 2.
Physical quantities are normalized in terms of
the speed of light $c$, the Schwarzschild radius $r_{\rm g}$, and
the Eddington luminosity $L_{\rm E}$
[$=4\pi cGM/(\kappa_{\rm abs}+\kappa_{\rm sca})$];
the units of $F$ and $cP$ are $L_{\rm E}/(4\pi r_{\rm g}^2)$.

\begin{figure}
  \begin{center}
  \FigureFile(80mm,80mm){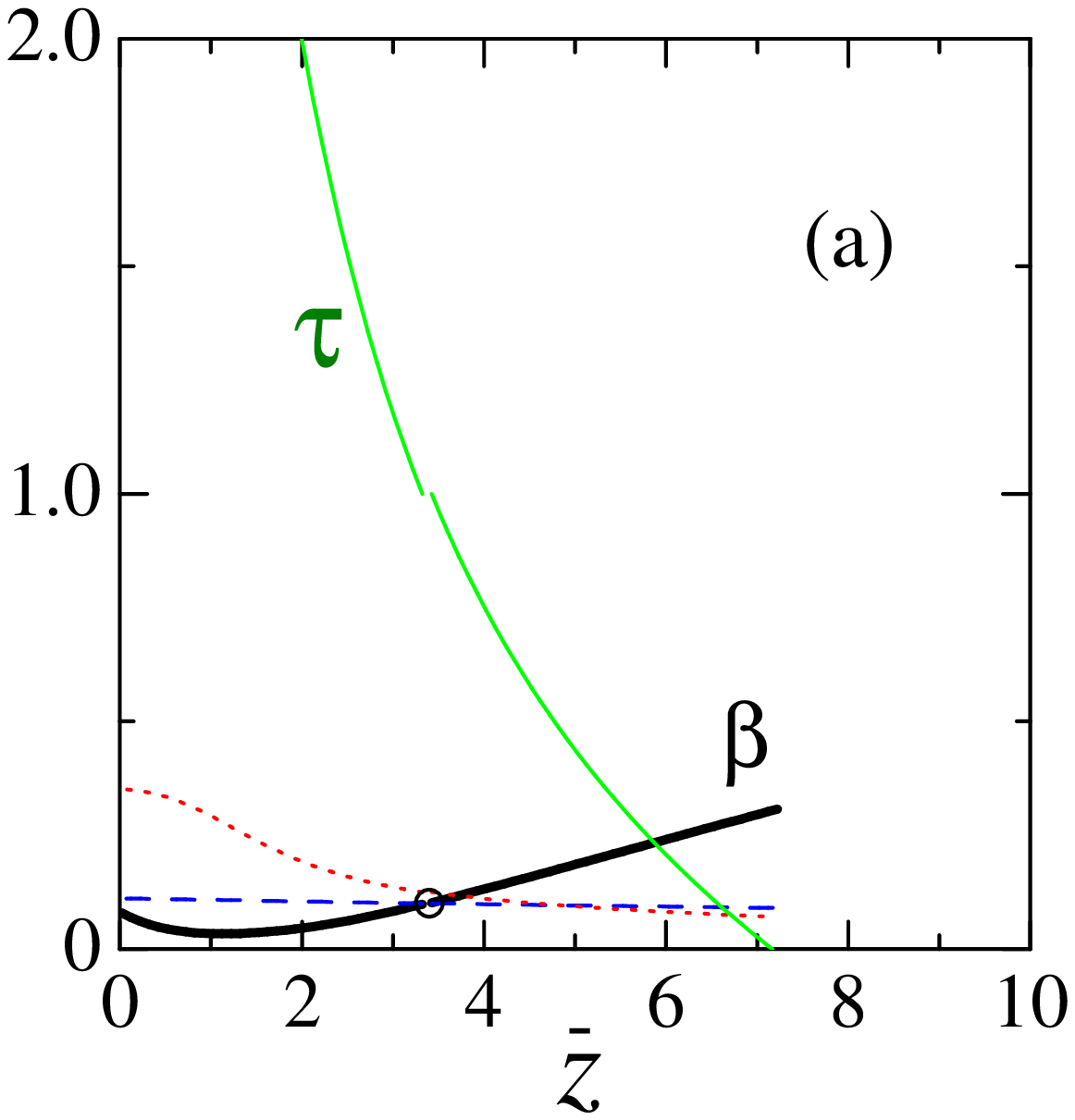}
  \FigureFile(80mm,80mm){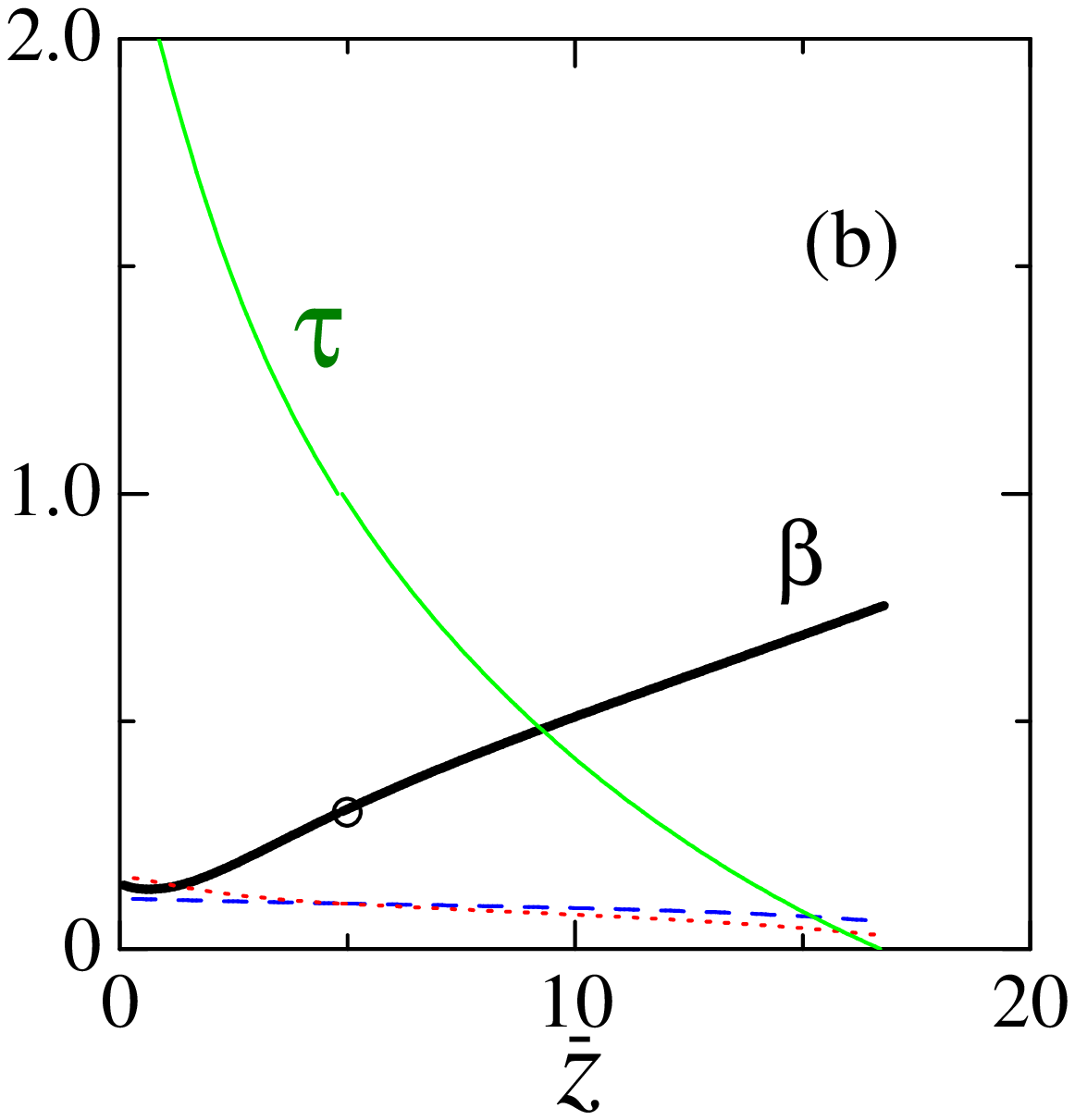}
  \FigureFile(80mm,80mm){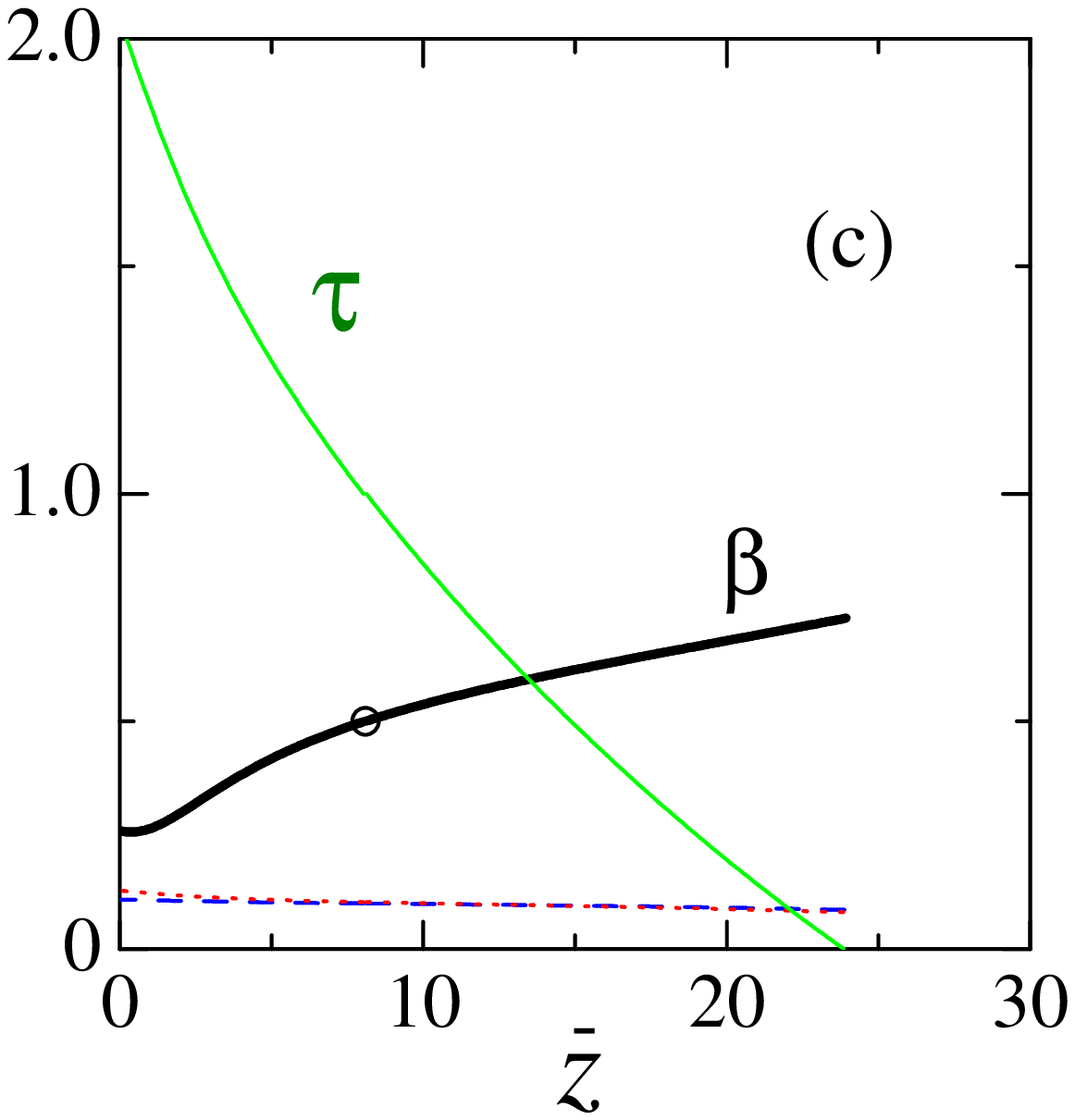}
  \end{center}
\caption{
Flow velocity $\beta$ (thick solid curve),
radiative flux $\hat{F}$ (dashed one),
radiation pressure $\hat{P}$ (dotted one),
and optical depth $\tau$ (solid curve),
as a function of the height $\hat{z}$.
The parameters are $\hat{r}=3$, $\delta=1$,
$\tau_{\rm c}=1$, $\hat{F}_{\rm c}=0.1$, and $\hat{J}=0.1$.
Furthermore, for each panel,
(a) $\hat{c}_{\rm T}=0.1$, then, $\hat{z}_{\rm c}=3.37$, $\hat{P}_{\rm c}=0.125$
    and $\beta_{\rm s}=0.3034$,
(b) $\hat{c}_{\rm T}=0.3$, then, $\hat{z}_{\rm c}=5.70$, $\hat{P}_{\rm c}=0.108$
    and $\beta_{\rm s}=0.5644$,
(c) $\hat{c}_{\rm T}=0.5$, then, $\hat{z}_{\rm c}=8.07$, $\hat{P}_{\rm c}=0.103$
    and $\beta_{\rm s}=0.7268$.
Critical points are marked by open circles.
}
\end{figure}

In figure 2
we show the flow velocity $\beta$ (thick solid curve),
the radiative flux $\hat{F}$ (dashed one),
the radiation pressure $\hat{P}$ (dotted one),
and the optical depth $\tau$ (solid curve),
as a function of the height $\hat{z}$
for $\hat{r}=3$, $\delta=1$, $\tau_{\rm c}=1$, $\hat{F}_{\rm c}=1$, 
$\hat{J}=0.1$,
and $\hat{c}_{\rm T}=$0.1, 0.3, 0.5.
Critical points are marked by open circles.

As is seen in, e.g., figure 2a,
the wind velocity slightly decreases at first,
because of the existence of the center-type critical point,
then increases to pass through the saddle-type critical point,
and finally reaches the final speed at the wind top of $\tau=0$.
The radiative flux $F$ slightly decreases,
as the height increases.
In the relativistic radiation flow,
the radiative flux does not conserve, but decreases,
since the radiation field acts to accelerate the gas.
As for the effect of the gas pressure,
the velocity fields entirely increase
as the isothermal sound speed increases.

The effects of the radiation force and gas pressure
on the acceleration of winds are summarized in figures 3 and 4.

In figure 3
we show the wind final velocity $\beta_{\rm s}$ (thick solid curve),
the heights, $\hat{z}_{\rm s}$ and $\hat{z}_{\rm c}$,
 of the wind top and the critical point (dashed ones),
and the radiation pressure $\hat{P}_{\rm c}$ 
at the critical point (dotted one),
as a function of the radiative flux $\hat{F}_{\rm c}$ at the critical point.
The other parameters are fixed as $\hat{r}=3$, $\delta=1$,
$\tau_{\rm c}=1$, $\hat{J}=0.1$, and $\hat{c}_{\rm T}=0.3$.

As is seen in figure 3,
the wind final velocity increases with $\hat{F}_{\rm c}$.
The Lorentz factor $\gamma_{\rm s}$ 
of the wind final velocity is well fitted by
\begin{equation}
   \gamma_{\rm s} \sim 1.07 + 1.50 \hat{F}_{\rm c}.
\end{equation}
Roughly speaking, this is understood as follows.
In the present treatment under special relativity
and pseudo-Newtonian potential,
there is no rigorous energy integral.
However, the pseudo total energy along the wind,
\begin{equation}
   Jc^2 \frac{\varepsilon+p}{\rho c^2} \gamma
   \left[ 1 - \frac{r_{\rm g}}{2(R-r_{\rm g})} \right] + F,
\end{equation}
is approximately conserved within the error of 10\%.
If we denote the gravitational potential by $\phi$,
this pseudo energy conservation is written in the non-dimensional form as
\begin{equation}
   \frac{\hat{J}}{\delta^2} \gamma (1+\hat{\phi}) + \hat{F} = {\rm const.},
\end{equation}
or
\begin{equation}
   \left. \frac{\hat{J}}{\delta^2} \gamma (1+\hat{\phi}) + \hat{F} \right|_{\rm s}=
   \left. \frac{\hat{J}}{\delta^2} \gamma (1+\hat{\phi}) + \hat{F} \right|_{\rm c}.
\end{equation}
Hence, in general $\gamma_{\rm s}$ linearly depends on $\hat{F}_{\rm c}$.

In figure 4
we show the wind velocity $\beta_{\rm s}$ (thick solid curve),
the heights, $\hat{z}_{\rm s}$ and $\hat{z}_{\rm c}$,
 of the wind top and the critical point (dashed ones),
and the radiation pressure $\hat{P}_{\rm c}$ 
at the critical point (dotted one),
as a function of the isothermal sound speed $\hat{c}_{\rm T}$.
The other parameters are fixed as $\hat{r}=3$, $\delta=1$,
$\tau_{\rm c}=1$, $\hat{J}=0.1$, and $\hat{F}_{\rm c}=0.1$.

As is seen in figure 4,
the wind final velocity increases with $\hat{c}_{\rm T}$.
The Lorentz factor $\gamma_{\rm s}$ is now fitted by
\begin{equation}
   \gamma_{\rm s} \sim 1 + 0.3 \hat{c}_{\rm T} + 1.3 \hat{c}_{\rm T}^2.
\end{equation}
This is also understood by a rough energy conservation
discussed above.
In this case, $\gamma_{\rm s}$ linearly depends on $\gamma_{\rm c}$,
and $\gamma_{\rm c}=1/\sqrt{1-\hat{c}_{\rm T}^2}$,
therefore there appears $\hat{c}_{\rm T}^2$ term.

\begin{figure}
  \begin{center}
  \FigureFile(80mm,80mm){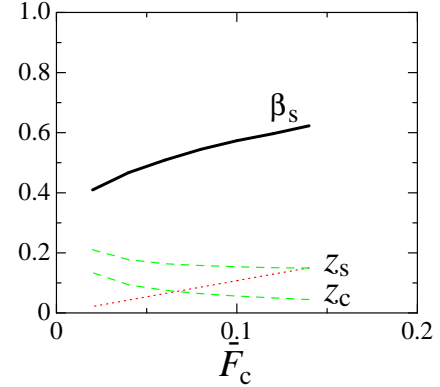}
  \end{center}
\caption{
Wind final velocity $\beta_{\rm s}$ at the wind top (thick solid curve),
heights, $\hat{z}_{\rm s}$ and $\hat{z}_{\rm c}$,
 of the wind top and the critical point (dashed ones),
and radiation pressure $\hat{P}_{\rm c}$ 
at the critical point (dotted one),
as a function of the radiative flux $\hat{F}_{\rm c}$ at the critical point.
For heights, $\hat{z}/100$ is plotted.
The parameters are $\hat{r}=3$, $\delta=1$,
$\tau_{\rm c}=1$, $\hat{J}=0.1$, and $\hat{c}_{\rm T}=0.3$.
}
\end{figure}

\begin{figure}
  \begin{center}
  \FigureFile(80mm,80mm){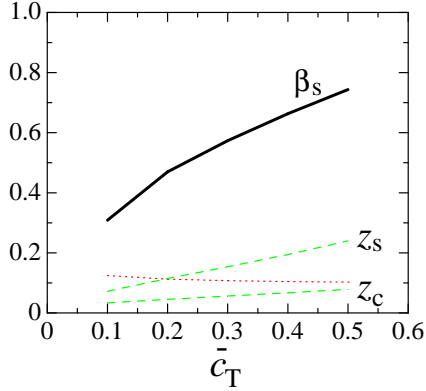}
  \end{center}
\caption{
Wind final velocity $\beta_{\rm s}$ at the wind top (thick solid curve),
heights, $\hat{z}_{\rm s}$ and $\hat{z}_{\rm c}$,
 of the wind top and the critical point (dashed ones),
and radiation pressure $\hat{P}_{\rm c}$ 
at the critical point (dotted one),
as a function of the isothermal sound speed $\hat{c}_{\rm T}$.
For heights, $\hat{z}/100$ is plotted.
The parameters are $\hat{r}=3$, $\delta=1$,
$\tau_{\rm c}=1$, $\hat{J}=0.1$, and $\hat{F}_{\rm c}=0.1$.
}
\end{figure}

\begin{figure}
  \begin{center}
  \FigureFile(80mm,80mm){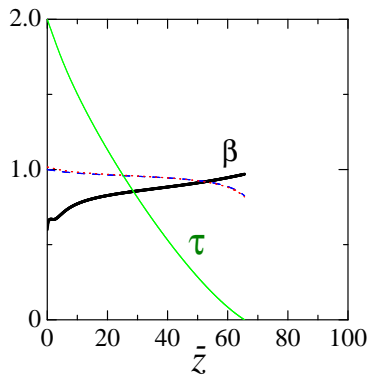}
  \end{center}
\caption{
Flow velocity $\beta$ (thick solid curve),
radiative flux $\hat{F}$ (dashed one),
radiation pressure $\hat{P}$ (dotted one),
and optical depth $\tau$ (solid curve),
as a function of the height $\hat{z}$.
The parameters are $\hat{r}=3$, $\delta=1$,
$\tau_{\rm 0}=2$, $\hat{F}_{\rm 0}=1$, $\beta_0=0.6$, $\hat{c}_{\rm T}=0.57$
and $\hat{J}=0.1$.
}
\end{figure}

In addition to transonic solutions,
an example of supersonic solutions
is shown in figure 5.
In figure 5
we show the flow velocity $\beta$ (thick solid curve),
the radiative flux $\hat{F}$ (dashed one),
the radiation pressure $\hat{P}$ (dotted one),
and the optical depth $\tau$ (solid curve),
as a function of the height $z$
for $\hat{r}=3$, $\delta=1$,
$\tau_{\rm 0}=2$, $\hat{F}_{\rm 0}=1$, $\beta_0=0.6$, $\hat{c}_{\rm T}=0.57$,
and $\hat{J}=0.1$,
where the subscript 0 means the values at $z=0$.

In this example of supersonic winds,
the wind final speed is almost the speed of light,
since both the initial flux and isothermal sound speed are large.

\section{Concluding Remarks}

In this paper 
we have examined the relativistic radiation hydrodynamical winds
 from a luminous accretion disk
in the relativistic regime of $(v/c)^2$.
The wind is assumed to be steady, vertical, and isothermal.
Using a velocity-dependent variable Eddington factor,
the basic equations can be numerically solved
without meeting the pathological singular point
at $v=c/\sqrt{3}$ to reach the relativistic regime.

For less luminous cases,
vertical disk winds are transonic types
passing through saddle-type critical points,
and the wind final speeds are 0.4--0.8$~c$ for typical parameters.
For luminous cases, on the other hand,
disk winds become supersonic types
without passing through any critical points,
and the wind final speeds becomes on the order of $c$.
The boundary between the transonic and supersonic types is located
at around 
$\hat{F}_{\rm c} \sim 0.1 (\varepsilon+p)/(\rho c^2)/\gamma_{\rm c}$.

In usual standard accretion disks,
the local luminosity is sub-Eddington.
Hence, for proton-electron normal plasmas
the above condition is safficiently fulfilled,
and transonic winds driven by radiation and gas pressures
would blow off.
For electron-positron pair plasmas, on the other hand,
the local luminosity becomes super-Eddington,
and pair winds would supersonically blow off.

The maximum attainable velocity of the wind
can be roughly estimated by the energy conservation.
In the Schwarzschild space-time, without any energy source
except for gravity,
the total energy of the radiation hydrodynamical flow
along the streamline, 
\begin{equation}
   Jc^2 \frac{\varepsilon+p}{\rho c^2}
   \gamma \sqrt{g_{00}} + g_{00} F
   = \dot{E} ~({\rm const.}),
\end{equation}
where $g_{00} = \sqrt{1-r_{\rm g}/R}$,
is conserved.
If all of the thermal energy of the gas
and the radiative energy in the initial state
is converted to the bulk energy of the gas,
the final velocity would become its upper limit:
\begin{equation}
   Jc^2 \gamma_\infty=
   \left. Jc^2 \frac{\varepsilon+p}{\rho c^2}
   \gamma \sqrt{g_{00}} + g_{00} F \right|_{z=0}.
\end{equation}
If, further, the gas is at rest in the initial state
and in the virial state,
$[(\varepsilon+p)/(\rho c^2)]\sqrt{g_{00}} |_{z=0} = 1$,
then we finally have the maximum attainable velocity as
\begin{equation}
   \gamma_\infty = 1+\frac{g_{00}F |_0}{Jc^2}
                 = 1+\left( 1-\frac{1}{\hat{r}} \right)
                     \frac{\hat{F}_0}{\hat{J}}.
\end{equation}
For example,
in the present typical case of
$\hat{r}=3$, $\hat{F} \sim 0.1$, $\hat{J}=0.1$,
we have $\gamma_\infty=1.666$, or $\beta_\infty=0.8$.

Hence, from the view point of energetics,
the final velocity becomes higher and higher,
as the ratio $\hat{F}_0/\hat{J}$ becomes large;
the radiative flux is so high or
the mass-loss rate is so low.
In other words,
if the small amount of the gas gains
the large amount of the radiation energy,
the final bulk velocity becomes high.
In general, however, it may be difficult
for such a re-distribution of energy to take place,
and the ratio $\hat{F}_0/\hat{J}$
would be on the order of unity.

Thus, we conclude that
the maximum attainable velocity
of accretion disk winds
emanating from the disk inner region
would be $0.8~c$ or so,
as long as it consists of normal plasmas.
This conclusion is consistent with
the current theoretical works refered in the introduction,
those explain mildly relativistic jets
of $0.26~c$ ($\gamma=1.04$) in SS~433 to highly relativistic jets
of $0.92~c$ ($\gamma=2.55$) in several microquasars.

For ultra-relativistic jets of $0.99~c$ ($\gamma=10$) 
supposed in several active galactic nuclei or
extremely relativistic jets of $0.9999~c$ ($\gamma=100$)
expected to gamma-ray bursts,
it may be necessary some other processes,
including
an extra-ordinary re-distribution of energy,
pair dominant plasmas,
an energy deposition from other energy sources,
such as a nuclear one via neutrino,
and so on.

\vspace*{1pc}

This work has been supported in part
by a Grant-in-Aid for the Scientific Research (18540240 J.F.) 
of the Ministry of Education, Culture, Sports, Science and Technology.


\end{document}